# Band degeneracy, resonant level formation and low thermal conductivity in dilute In and Ga co-doped thermoelectric compound SnTe

*Gaurav Jamwal[1#], Ankit Kumar[2], Mohd. Warish[1], Shruti Chakravarty[2], Saravanan Muthiah[3], Asokan Kandasami[4,5], and Asad Niazi[1\*]*

[1] Department of Physics, Jamia Millia Islamia, New Delhi – 110025, India
[2] Department of Physics, IISER Pune, Dr. Homi Bhabha Road, Pune – 411008, India
[3] CSIR-National Physical Laboratory, Dr. K.S. Krishnan Marg, New Delhi – 110012, India
[4] Inter-University Accelerator Centre, Aruna Asaf Ali Marg, New Delhi – 110067, India
[5] Department of Physics & Center for Interdisciplinary Research, University of Petroleum and Energy Studies (UPES), Dehradun – 248007, India
[#]Email: gaurav.jamwal33@gmail.com
[*]Email: aniazi@jmi.ac.in

**Abstract**

We report the effect of co-doping of In and Ga at low concentrations on the structural, electronic, and thermoelectric properties of SnTe based compositions $Sn_{1.03-2x}In_xGa_xTe$ (x = 0, 0.01, 0.02, 0.04) prepared by the solid-state route and spark plasma sintering (SPS). All compositions formed in the *fcc* structure (Fm$\overline{3}$m) with no other impurity phase. The optical band gap increased with the co-doping, indicative of band convergence effects. First principle electronic structure calculations showed band convergence and the formation of resonant levels, due to Ga and In doping respectively. The carrier concentration increased on hole-doping by In and Ga ions while carrier mobility decreased due to impurity scattering. The resistivity increased with temperature, indicative of the degenerate semiconducting character of the compounds. The Seebeck coefficient of the doped samples increased linearly with temperature, reaching 85 - 95 $\mu$V/K at 783 K. Thermal conductivity decreased sharply with co-doping, and the lattice thermal conductivity dropped to 0.42 Wm$^{-1}$K$^{-1}$ above 750 K. The enhanced power factor and low lattice thermal conductivity on doping resulted in a maximum figure of merit $ZT = 0.34$ at 773 K, twice that of the pristine SnTe.

# 1. INTRODUCTION

The search for alternative and renewable sources of energy has become increasingly urgent due to the ever-growing energy demand and environmental issues caused by fossil fuels. Thermoelectric (TE) materials have drawn worldwide attention as these can directly and reversibly convert heat into electricity, providing a viable sustainable, reliable and environmentally friendly alternative solution to the current energy crisis [1,2].

The viability of any device depends upon several factors such as ease of production, cost and efficiency. The conversion efficiency of a thermoelectric device is governed by the thermoelectric figure of merit, defined as $ZT = (S^2\sigma T)/\kappa$, where $S$, $\sigma$, $T$ and $\kappa$ are the Seebeck coefficient, electrical conductivity, working temperature and thermal conductivity (which is the sum of lattice thermal conductivity $\kappa_L$ and electron thermal conductivity $\kappa_E$) respectively. To optimize the thermoelectric performance of a material, a large $S$, high electrical conductivity and low thermal conductivity should ideally be simultaneously obtained [3,4]. However, these parameters are interdependent, making it very difficult to manipulate the individual properties independently. The strategies that can enhance the power factor $S^2\sigma$ include the improvement of $S$ through convergence of degenerate valleys of electronic bands [5,6], carrier engineering [7], and the introduction of resonance levels in the electronic bands [8,9]. Low thermal conductivity has been achieved through phonon scattering by point defects [10,11], nanoscale precipitates, nano/meso-architectures and intrinsic anharmonic phonons [12–14]. Some examples of important TE semiconductor materials are $Bi_2Te_3$ for refrigeration near room temperature, filled skutterudites ($CoSb_3$), PbTe for mid-temperature power generation (700 – 900°C) and SiGe for high temperature thermoelectric applications (> 1000°C) [15].

Lead chalcogenides have long been investigated as promising thermoelectric materials in the intermediate temperature range. Lead telluride (PbTe) has been extensively investigated as an efficient thermoelectric material for space and military applications since the 1950s [16]. Numerous advances have been made in PbTe-based compounds, with a high $ZT \sim 2.2$ at 915 K obtained by band engineering techniques [17,18]. However, environmental concerns due to the toxicity of Pb prevents their large-scale applications in thermoelectric power generation [18]. Therefore, there is a need to explore alternative materials with higher efficiency for waste heat recovery.

Recently, tin telluride (SnTe), a lead-free compound has been received much attention as a possible substitute for PbTe and many research groups have reported high thermoelectric properties of SnSe single crystals ($ZT$ = 2.6 at 900 K) [19]. Tin telluride (SnTe) exhibits rock salt crystal structure and electronic band structure similar to lead telluride (PbTe). Pure SnTe contains intrinsic vacancies at the Sn site, making it a heavily doped p-type semiconductor with a very high hole concentration ($\sim 10^{21}$ cm$^{-3}$), leading to large electrical and thermal conductivities [20]. Consequently, it exhibits a low Seebeck coefficient with a low figure of merit ($ZT \sim 0.4$), leading to poor thermoelectric performance. SnTe also has a larger separation ($\Delta E_v \sim 0.35$ eV) as compared to PbTe ($\Delta E_v \sim 0.17$ eV) between the higher lying light hole valence band at the L point and the lower lying heavy hole valence band at the Σ point [18,20–23]. This limits the hole transport to within a single band, giving a low *S*. Finally, SnTe has also a higher thermal conductivity ($\sim$ 3.5 Wm$^{-1}$K$^{-1}$) as compared to PbTe ($\sim$ 1.5 Wm$^{-1}$K$^{-1}$) since Sn is lighter than Pb [22,23]. Therefore, it is currently a significant challenge to improve the thermoelectric performance of SnTe to the same level as PbTe.

Several efforts have been made to resolve these problems in pristine SnTe via band structure convergence and structural manipulation at the micro and nano scales. Effective methods to suppress the excess (hole) carrier concentration to an appropriate level (∼$10^{19}$ cm$^{-3}$) are either by electron donor doping or by self -compensation of Sn. Several attempts have been carried out to modify the valence band structure of SnTe. Doping of Cd, Mg, Hg, Mn and Ca in SnTe provides efficient valence band convergence, which give rises to a significant improvement of the $S$ of SnTe [4,10,24–27]. Indium doping of SnTe has been shown to improve its $S$ due to the formation of resonance states in the valence band which results in a $ZT \sim 1.1$ at 873 K [9]. A reduction of lattice thermal conductivity has been reported in Ga doped SnTe via phonon scattering, resulting in a $ZT \sim 1$ at 873 K [28]. Co-doping of In-Cd, In-Ag and Ge-Mn has been carried out for the enhancement of the thermoelectric properties of SnTe via band engineering [29–31]. SnTe, containing excess Sn to compensate for intrinsic vacancies, has been investigated by co-doping with In-Cd and In-Mg [32,33]. Thus, different strategies via resonant level, band convergence and phonon scattering, have been employed to improve thermoelectric properties of pure SnTe. In this work, we report the enhancement of the thermoelectric performance of SnTe compositions containing excess Sn, by substitution of Sn through the co-doping of In and Ga.

## 2. EXPERIMENTAL SECTION

**2.1. Synthesis.** The chemicals: tin (99.9%, Alfa Aesar), tellurium (99.997%, Sigma Aldrich), indium ( 99.99%, Alfa Aesar), gallium (99.99%, Sigma Aldrich) were used for the synthesis without further purification. The Sn$_{1.03-2x}$In$_x$Ga$_x$Te (x = 0, 0.01, 0.02, 0.04) samples were synthesized by mixing appropriate ratios of high purity starting materials of Sn, In, Ga and Te in evacuated, sealed quartz ampoules under a vacuum below $10^{-5}$ Pa by the high temperature solid state route. The samples were heated slowly in a programmable box furnace to 1173 K, kept for 12 hours, and allowed to cool down to room temperature. The ingots obtained were ground in an agate mortar and the powders were loaded into graphite dies for spark plasma sintering (SPS). Highly dense circular disks were obtained by SPS at 850 K for 5 min under 50 MPa pressure in a high-vacuum environment. The density ($\rho_g$) of the samples as measured by the Archimedes method was determined to be > 99% of the maximum (Table S1).

**2.2. Physical Characterization.** Powder X-ray diffraction for all samples was recorded using Cu $K_\alpha$ radiation ($\lambda$ = 1.5406 Å) in the range 2θ = 20º - 80º on a Panalytical diffractometer (Empyrean PANalytical). The XRD data were analyzed by Rietveld refinement using Fullprof software. The Raman measurements were performed using a Renishaw (Invia Reflex) system with a 514 nm excitation laser in the wave number range from 52 to 300 cm$^{-1}$ at room temperature. To find the optical band gap, diffuse reflectance measurement has been performed at room temperature using an FTIR Bruker (VORTEX 70) spectrometer in a frequency range 4000 - 400 cm$^{-1}$. The morphology of compounds was recorded through field emission SEM (Tescan MIRA FE-SEM Oxford INCA PentaFETx3 EDS) and TEM (JEM F-200 Multipurpose Electron Microscope). The thermal stability with thermogravimetric (TGA) was studied using Thermal Gravimetric Analyzer STA 6000, EXSTAR in the temperature range from RT to 1080 K at a heating rate of 283 K min$^{-1}$ under a nitrogen atmosphere. X-ray photoelectron spectroscopy (XPS: SPECS, Germany) equipped with monochromatic Al $K_\alpha$ X-ray source (h$\nu$ = 1486.61 eV, X-ray energy: 13KV, 100 W) was used to investigate the binding energy and chemical bonding state. A Hall effect measurement system (Ecopia HMS-3000) was used to study the room temperature mobility, type of charge carrier, and carrier concentration. The electrical resistivity ($\rho$) and Seebeck coefficient ($S$) were measured simultaneously using a LINSEIS-LSR-3 instrument under helium atmosphere from room temperature (~ 300 K) to 785 K. The thermal diffusivity coefficient ($D$) was measured by the laser flash technique using a LINSEIS-LFA-1000 instrument. The density ($\rho_g$) of all samples was measured by the Archimedes method. The specific heat ($C_p$), of the samples was calculated using the Dulong-Petit law. The total thermal conductivity ($\kappa$) of the samples was calculated from the above values of $D$, $C_p$ and $\rho_g$. The power factor and figure of merit were estimated by using the measured parameters.

**2.3. Computational Details**

First-principle calculations were done using Quantum Espresso (QE) [34,35] density functional theory (DFT) code with the generalized gradient approximation Perdew-Burke-Ernzerhof (GGA-PBE) exchange correlation functional to describe the electron-electron exchange and correlation interactions [36]. Ultrasoft pseudopotentials were used to describe electron-ion interactions [37]. A kinetic energy cutoff of 55 Ry was used for the plane wave basis and 550 Ry for the charge density expansion, ten times that of the wave function. The density of states calculation was done on a $\sqrt{2} \times \sqrt{2} \times 2$ supercell containing 16 atoms of Sn and Te each. Doping by In or Ga was incorporated by replacing one of the 16 Sn atoms. A 10 x 10 x 7 Monkhorst-Pack k-point mesh was used for Brillouin zone integrations [38]. A smearing width of 0.005 Ry was used to speed up the convergence. For the band structure calculations, a supercell of 3 x 3 x 3 was used to incorporate dilute doping of In and Ga.

**3. RESULTS AND DISCUSSION**

**Powder XRD Diffraction**

The powder XRD diffraction patterns for the pure and samples $Sn_{1.03-2x}In_xGa_xTe$ alloys with x = 0, 0.01, 0.02, 0.04) are shown in figure 1(a). The patterns could be indexed to the cubic rock-salt NaCl-type structure with space group $Fm\bar{3}m$, indicating the formation of a single phase with no other impurity peaks observed. A small extraneous peak observed at 45° is due to the nickel sample holder, confirmed by repeating the XRD of the pure sample on another machine (Rigaku) with the powder coated on a glass slide (Supplemental Material (SM), Figure S1).

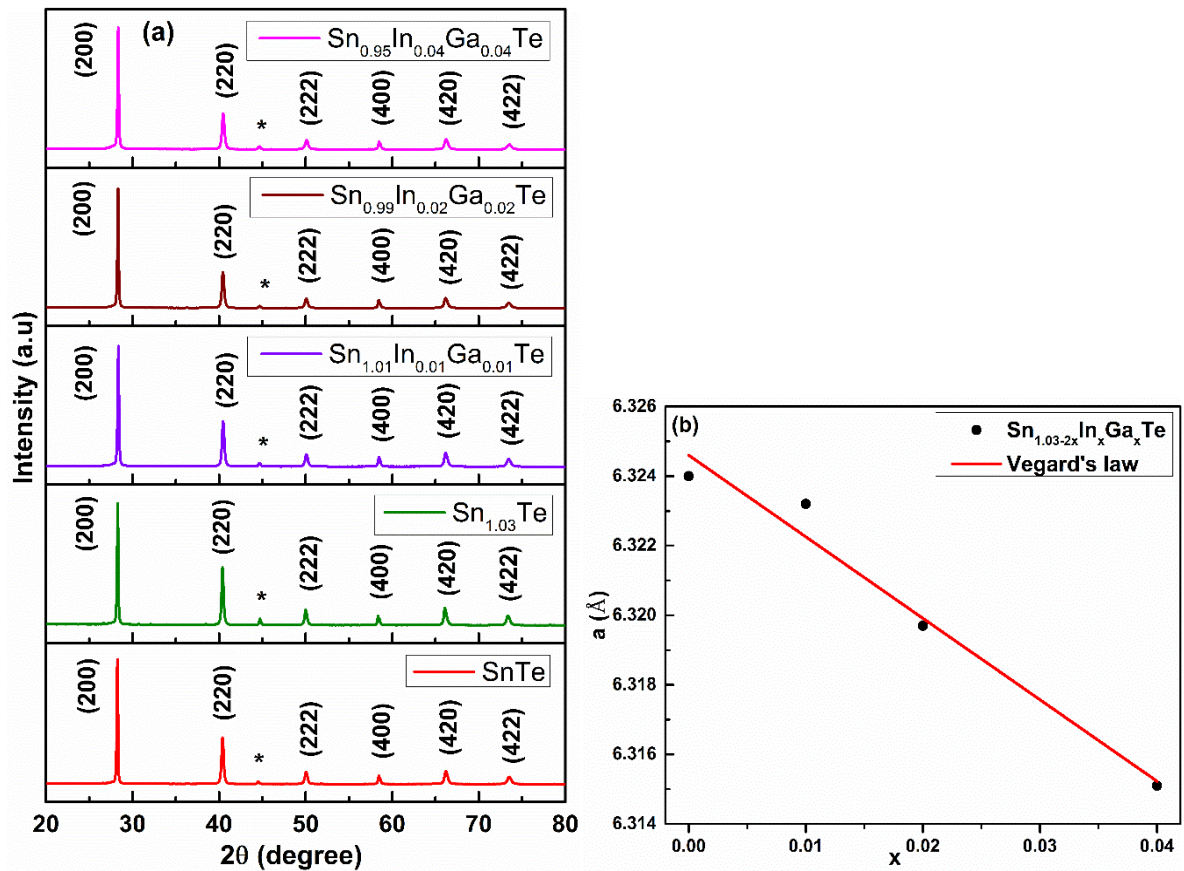

**Figure 1**. (a) Powder XRD patterns and (b) Lattice parameters, of polycrystalline SnTe and co-doped $Sn_{1.03-2x}In_xGa_xTe$ (x = 0, 0.01, 0.02, 0.04). The peak at 45º in (a) is due to the nickel sample holder.

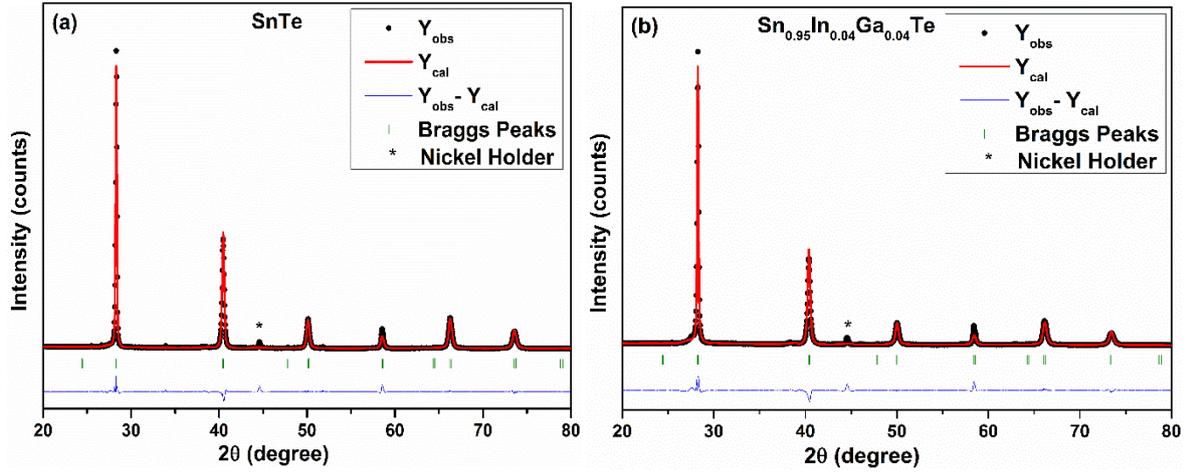

**Figure 2.** Rietveld refinements of SnTe and Sn$_{0.95}$In$_{0.04}$Ga$_{0.04}$Te using FullProf.

The powder XRD data of polycrystalline Sn$_{1.03-2x}$In$_x$Ga$_x$Te (x = 0, 0.01, 0.02, 0.04) samples were analyzed by Rietveld refinement using the FullProf Suite (Figure 2, S2) [39]. The results of Rietveld refinement are listed in Table I. The ionic radius of In$^{3+}$ (0.94 Å) and Ga$^{3+}$ (0.76 Å) are both smaller than that of Sn$^{2+}$ (1.18 Å) [40]. In accordance with Vegard's law, the lattice parameter decreases with increasing concentration of In-Ga co-doping, from 0.63245 nm for x = 0 to 0.63153 nm for x = 0.04, as shown in Figure 1(b).

Table I also lists the average crystallite size of the In-Ga co-doped SnTe compounds as estimated from the Debye-Scherrer formula (eq. 1):

$$D_c = \frac{k\lambda}{\beta_D \cos\theta} \quad (1)$$

Where, $D_c$ is the volume weighted crystallite size (nm), $k$ is the shape factor (0.9), $\lambda$ is wavelength of the X-rays ($\lambda$ = 1.54056 Å for Cu$k_\alpha$ radiation), $\theta$ is Bragg diffraction angle and $\beta_D$ is the broadening of the diffracted peak measured at FWHM (in radians).

**Table I**. Lattice parameters and grain size of cubic Sn$_{1.03-2x}$In$_x$Ga$_x$Te (x = 0, 0.01, 0.02, 0.04) obtained from powder XRD data.

| Composition | Lattice Parameter (Å) | Crystallite Size (nm) | $R_p$ | $R_{wp}$ | $\chi^2$ |
|---|---|---|---|---|---|
| SnTe | 6.3049 (2) | 36 | 7.5 | 11.5 | 7.8 |
| $Sn_{1.03}Te$ | 6.3240 (2) | 36 | 6.7 | 12.2 | 8.7 |
| $Sn_{1.01}In_{0.01}Ga_{0.01}Te$ | 6.3232 (3) | 33 | 7.6 | 11.9 | 9.5 |
| $Sn_{0.99}In_{0.02}Ga_{0.02}Te$ | 6.3197 (3) | 33 | 8.3 | 12.2 | 9.9 |
| $Sn_{0.95}In_{0.04}Ga_{0.04}Te$ | 6.3151 (3) | 32 | 8.4 | 12.4 | 9.6 |

**Morphological characterization**

To identify the morphology, composition, and phase formation, TEM studies were performed for the pure and co-doped (In, Ga) SnTe samples. Figure 3 shows high resolution TEM (HRTEM) images of SnTe and $Sn_{0.95}In_{0.04}Ga_{0.04}Te$. The interplanar d-spacing for some specific *hkl* directions was calculated from the Fourier-transformed (FT) patterns corresponding to selected SnTe crystallites in the HRTEM images. Figure 6(a, b) shows lattice fringes with the calculated d(*hkl*) values for the pure SnTe and $Sn_{0.95}In_{0.04}Ga_{0.04}Te$ compounds, which are in good correspondence with the parameters obtained from XRD.

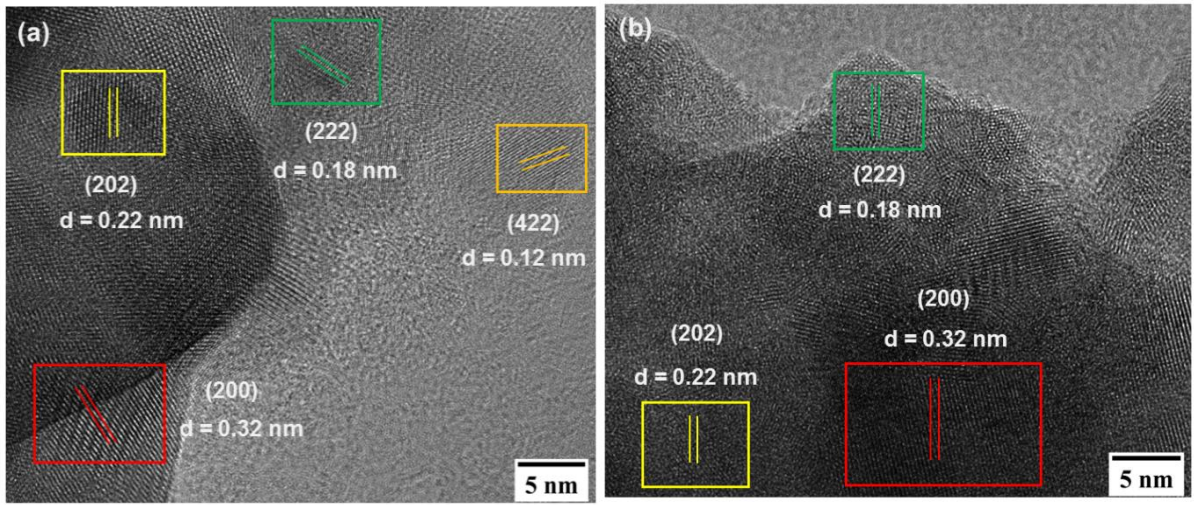

**Figure 3.** High-resolution (HR) TEM images of (a) pure SnTe and (b) $Sn_{0.95}In_{0.04}Ga_{0.04}Te$.

The SAED patterns for pure SnTe and $Sn_{0.95}In_{0.04}Ga_{0.04}Te$ are shown in Figure 4. The presence of sharp rings confirms the well-crystallized polycrystalline nature of compounds and corresponds to the (200), (220), (222) and (400) orientations of the cubic compound as shown in Figure 4(a,b). The results of scanning transmission electron microscopy (STEM), and energy dispersive X-ray spectroscopy (EDS) elemental mapping for pure SnTe and $Sn_{0.95}In_{0.04}Ga_{0.04}Te$ respectively are shown in Figure S3. The elemental mapping confirms that the spatial distribution of all the elements is homogeneous without any aggregation.

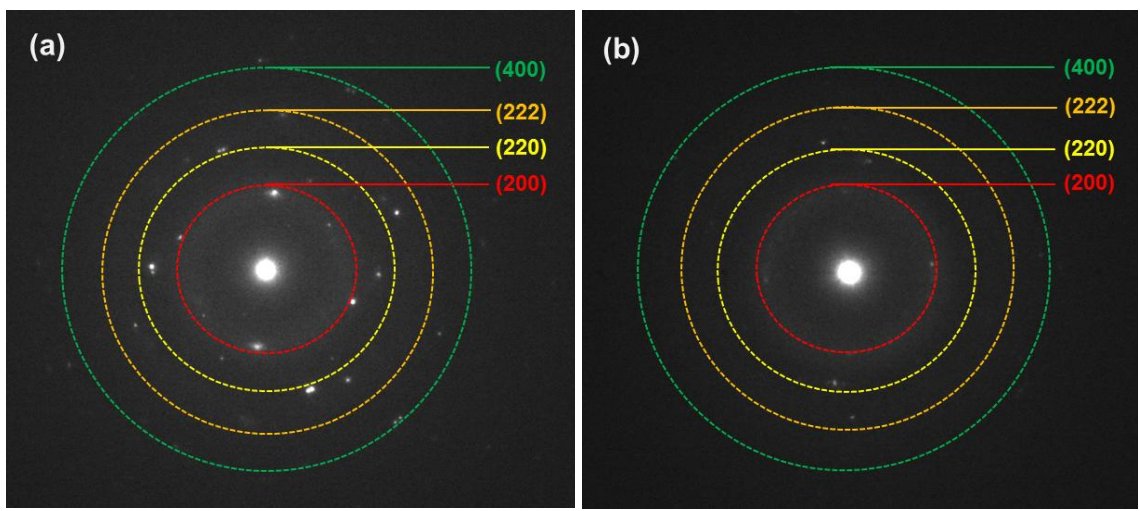

**Figure 4.** Selected area electron diffraction (SAED) pattern of pure (a) pure SnTe and (b) $Sn_{0.95}In_{0.04}Ga_{0.04}Te$ compound.

FESEM surface images of pure SnTe and $Sn_{1.03-2x}In_xGa_xTe$ sintered pellets (x = 0, 0.02, 0.04) are presented in Figure S4 and do not show any observable porosity. The In-Ga co-doping shows no significant effect on the microstructure.

Energy dispersive x-ray spectroscopy (EDS) measurements of sintered pellets of pure SnTe and $Sn_{1.03-2x}In_xGa_xTe$ (x = 0, 0.02, 0.04) were used to examine the chemical composition of samples (Figure S5). The EDS scans, recorded at 2μm resolution, show the constituent elements to be uniformly distributed in the specimens. Thus, from the FESEM and EDS measurements, it can be concluded that the sample matrix is homogeneous, smooth, and compact with a high relative density.

**Compositional analysis**

X-ray photoelectron spectroscopy (XPS) was performed to investigate the valence states of the elements and the chemical composition of the compounds. Figure 5 shows the XPS wide scan spectra of pure SnTe and In-Ga co-doped $Sn_{0.95}In_{0.04}Ga_{0.04}Te$. The core level spectra corresponding to the Sn, Te, In and Ga peaks are shown in Figure S6(a)-(e). The presence of carbon and oxygen was attributed to the oxidation of the sample upon atmospheric exposure while handling. The high resolution XPS spectral scans were deconvoluted using XPSpeak4.1 software [41] and are shown in figure 6(a)-(e). Figure S6(a) shows the Sn 3d core level peaks located at 489.2 and 497.8 eV corresponding to $3d_{5/2}$ and $3d_{3/2}$ levels respectively. The deconvolution results shown in figure 6(b) show two components for each peak, corresponding to $Sn^{2+}$ (487.9, 496.2 eV) and $Sn^{4+}$ (489.7, 498.1 eV) states.

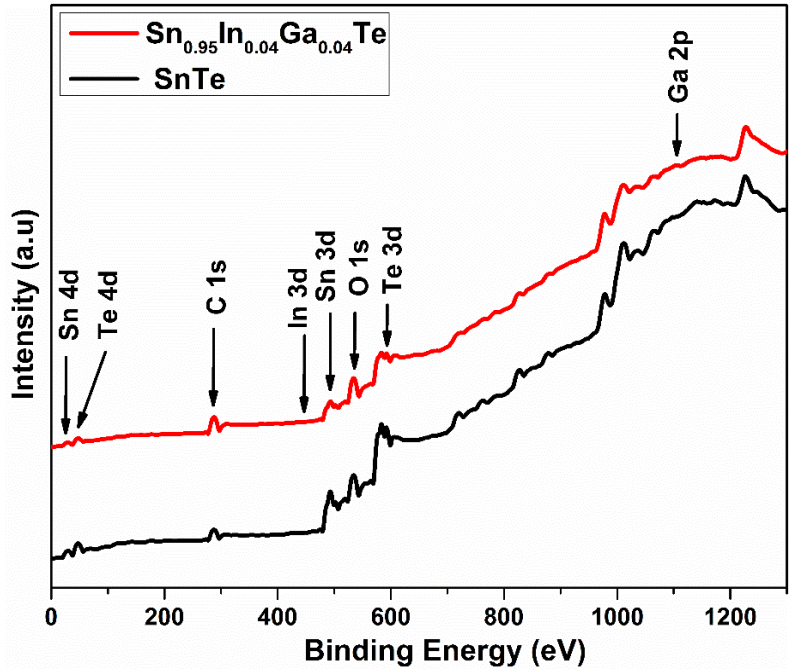

**Figure 5.** Full XPS spectra of pure SnTe and In-Ga co-doped Sn$_{0.95}$In$_{0.04}$Ga$_{0.04}$Te.

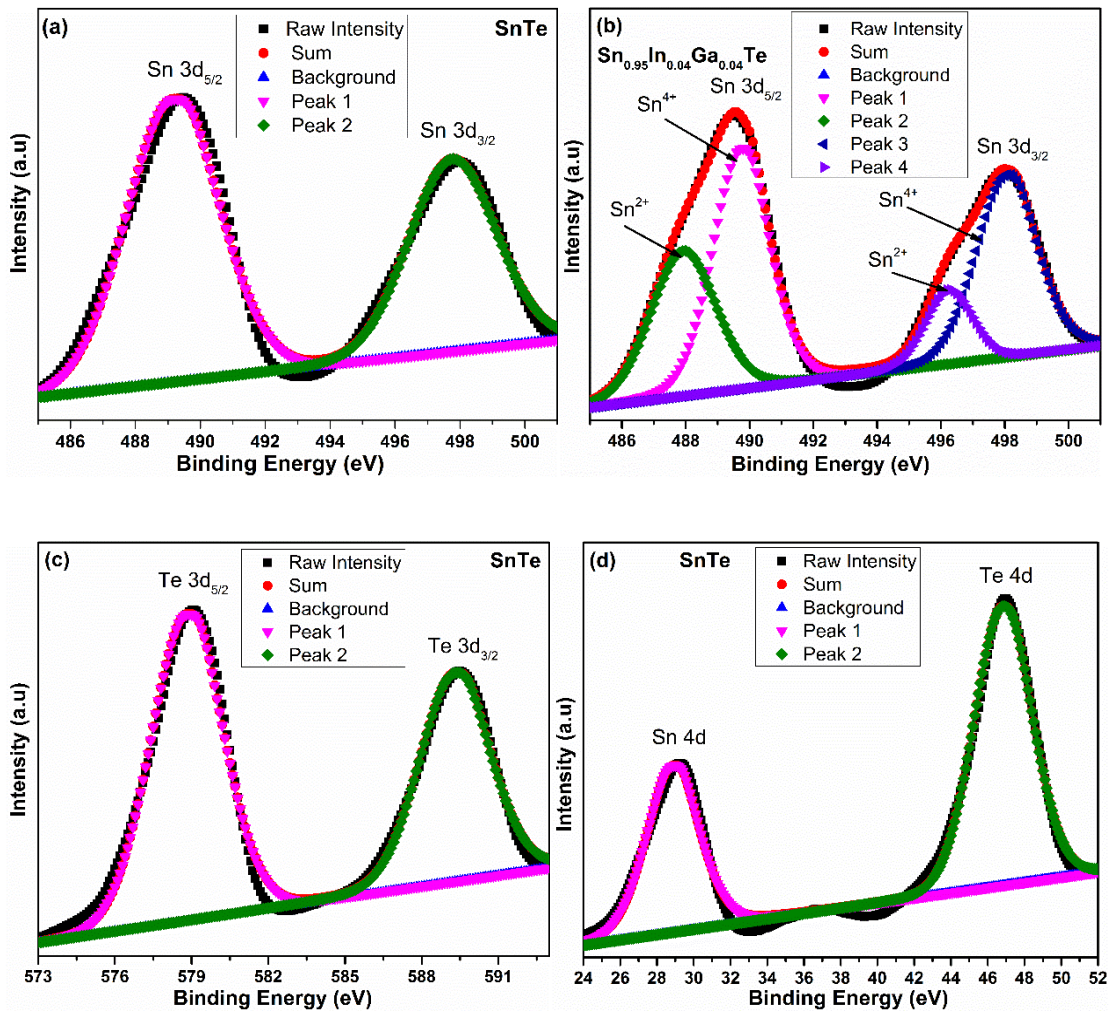

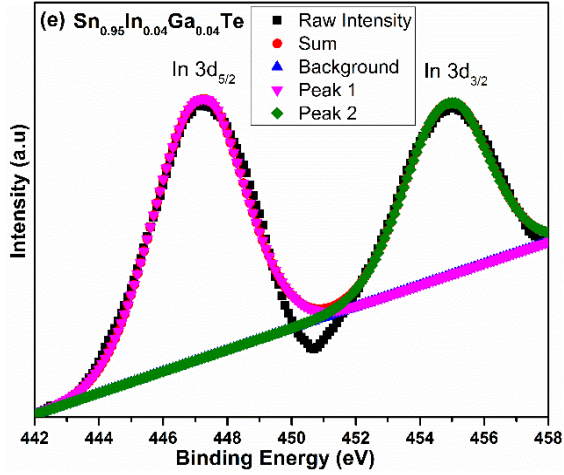

**Figure 6.** (a-e) Deconvoluted XPS spectra of pure SnTe and $Sn_{0.95}In_{0.04}Ga_{0.04}Te$ samples, showing contributions from different atomic states of the constituent elements, calculated using XPSpeak4.1 software.

However, there is no clear signature of elemental Sn (~ 485.2 eV) in the spectrum of the doped sample, which would be expected upon $In^{3+}$ and $Ga^{3+}$ co-doping. This can be explained by the presence of $Sn^{4+}$, as any elemental Sn present on/just beneath the sample surface would get oxidized to $Sn^{4+}$ upon exposure to atmospheric oxygen during handling. The peaks of the $Te^{2-}$ 3d spectrum (figure S6(b)) occur at binding energies of 579 eV ($3d_{5/2}$) and 589.3 eV ($3d_{3/2}$) respectively. Figure S6(c) shows the Sn 4d and Te 4d peaks at 28.9 eV and 46.9 eV respectively remain relatively unaffected upon co-doping. These results agree well with previously reported data on SnTe [42]. Indium was successfully doped into all the synthesized samples, as indicated by the two characteristic peaks of In $3d_{5/2}$ and In $3d_{3/2}$ located at 447.2 and 454.9 eV respectively as shown in figure S6(d). Figure S6(e) shows the Ga 2p core level peak at 1113 eV corresponding to the $2p_{3/2}$ core level. The observed XPS data are in good agreement with the calculated one. We conclude that In and Ga have been successfully co-doped into all synthesized samples as indicated by the XPS analysis.

**Raman spectroscopy**

The Raman spectra of pure SnTe and co-doped $Sn_{1.03-2x}In_xGa_xTe$ (x = 0, 0.01, 0.02, 0.04) samples are shown in Figure 7(a). Two dominant peaks are found in the frequency range of 52 cm$^{-1}$ to 300 cm$^{-1}$, which are assigned to the $A_1$ symmetry (optical phonon) and $E_{TO}$ (transverse optical phonon) breathing modes of tellurium in SnTe [43,44]. As the doping concentration increases, the $A_1$ and $E_{TO}$ modes shift towards higher wavenumbers. The $A_1$ (optical phonon) and $E_{TO}$ (transverse optical phonon) modes are shown schematically in Figure 7(b). The shift in these modes indicates the change in the chemical bond environment around Te upon doping.

The Raman shift occurs because the In-Ga co-doping breaks the center of inversion symmetry of SnTe and changes the bond strength (force constant) of Te-Sn (In-Ga) bonds. Using a classical vibrational model, the vibrational frequency is given by $f \sim (k_f/\mu_m)^{1/2}$ where $k_f$ is the force constant of a bond, and $\mu_m$ is the reduced mass [45]. Doping with a light element (In or Ga) at the Sn site decreases the reduced mass, as a result of which the Raman shift moves to a higher frequency. This is in agreement with the present results as well as previously reported studies [46].

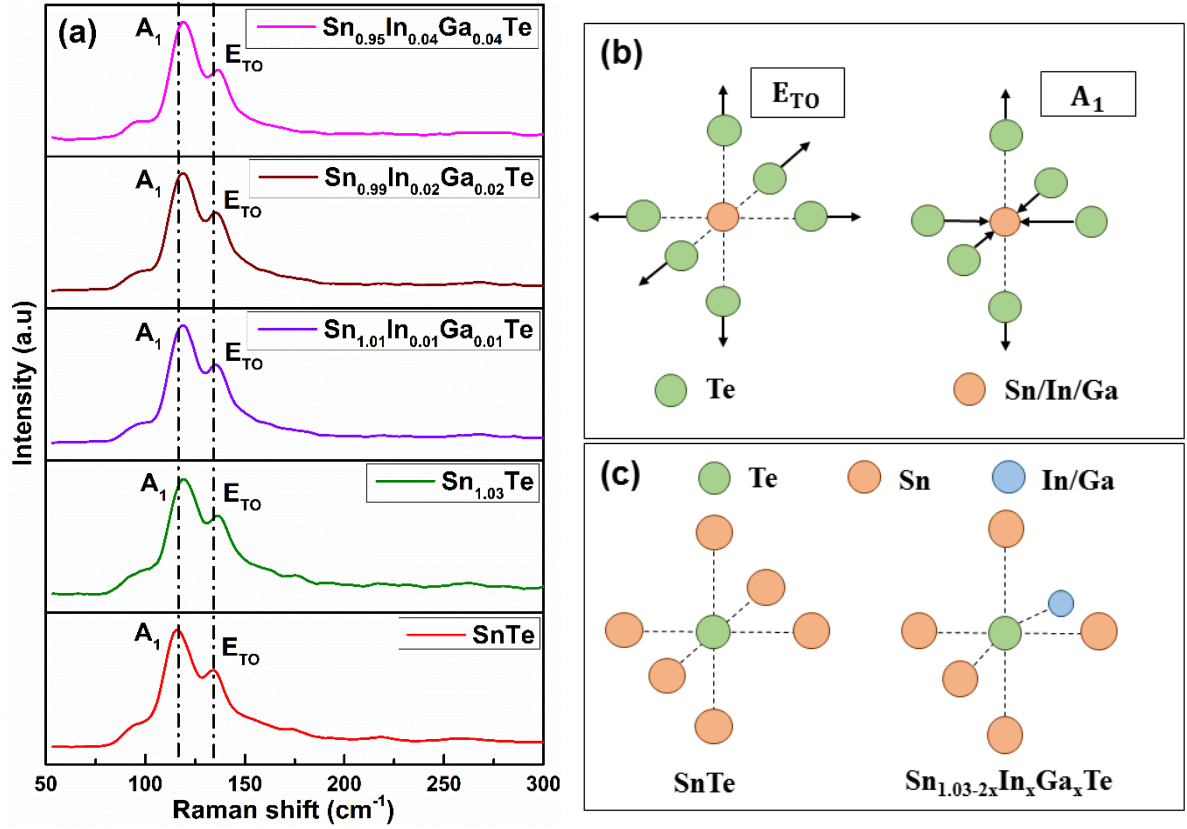

**Figure 7.** (a) Raman spectra of pure SnTe and co-doped $Sn_{1.03-2x}In_xGa_xTe$ (x = 0, 0.01, 0.02, 0.04), (b) Vibration schemes of $A_1$ and $E_{TO}$ modes for SnTe, (c) Schematic presentation of off-centering due to co-doping of In and Ga.

**Optical analysis**

The band gap of the $Sn_{1.03-2x}In_xGa_xTe$ (x = 0, 0.01, 0.02, 0.04) compositions was deduced from the measurement of diffuse reflectance FTIR spectra. The Kubelka-Munk (KM) formalism was used to transform the reflectance (R) into an absorption coefficient ($\alpha$), which is proportional to the KM function F(R), using the relation (eq. 2) [47,48]:

$$F(R) = \frac{(1-R)^2}{2R} = \frac{K}{S_{sc}} \qquad (2)$$

where, R is the measured absolute reflectance of sample (R = $R_{sample}$ / $R_{standard}$), K is the absorption coefficient, and $S_{sc}$ is the scattering coefficient for a given wavelength.

The band gap ($E_g$) and absorption coefficient ($\alpha$) of a direct band gap semiconductor are related by equation (3):

$$\alpha(h\nu) = A\,(h\nu - E_g)^{1/2} \quad (3)$$

where, $h\nu$ is the photon energy, and A is a material specific constant. By assuming that the absorption coefficient ($\alpha$) is proportional to the KM function F(R), the bandgap can be estimated from the plot of relation (eq. 4) [48,49]:

$$[F(R)(h\nu)]^2 = B\,(h\nu - E_g) \quad (4)$$

Figure 8 shows the plot of $[F(R)\,(h\nu)]^2$ as a function of photon energy for the representative composition $Sn_{1.03}Te$. The region of the curve from 0.28004 – 0.30013 eV (shown as dotted line) had a contribution from residual $CO_2$ inside the FTIR chamber. That data has been removed for clarity and replaced by the dotted curve, which does not affect the band gap determined from the extrapolated linear fit. The band gaps for the other compositions of $Sn_{1.03-2x}In_xGa_xTe$ (x = 0.01, 0.02 and 0.04) were determined in a similar manner. The direct band gap of SnTe (0.18 eV) could not be measured using diffuse reflectance IR spectroscopy because of strong interference from a large number of free carriers (excess holes).

The band gap values of $Sn_{1.03-2x}In_xGa_xTe$ are listed in Table II and show an increase with doping from $Sn_{1.03}Te \approx 0.21eV$ to $Sn_{0.95}In_{0.04}Ga_{0.04}Te \approx 0.25eV$. Previous studies have shown that doping leads to convergence of the light and heavy hole valence bands, along with an increase in the band gap [50,51]. This suggests that the increase in the band gap upon co-doping of (In, Ga) in SnTe may be associated with a reduction of energy separation between the light-hole and heavy-hole valence bands as indicated in the schematic diagram in Figure 9.

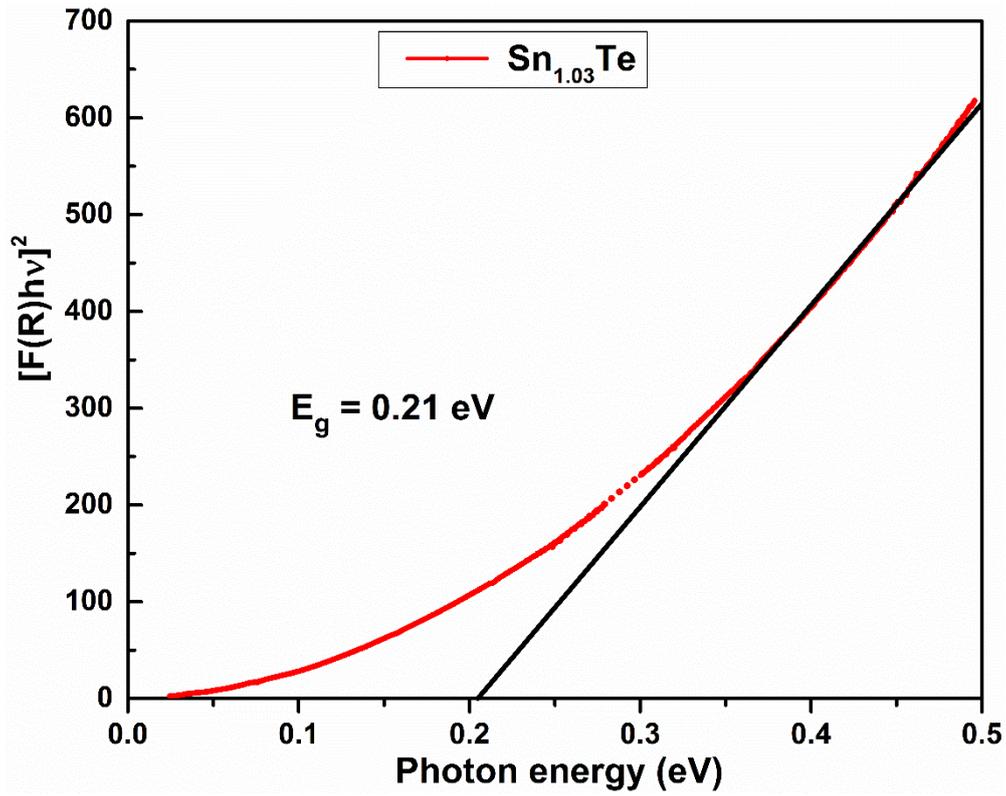

**Figure 8.** Kubelka-Munk transformed reflectance spectra of $Sn_{1.03}$Te.

**Table II.** Direct band gap of $Sn_{1.03-2x}In_xGa_xTe$ determined from FTIR measurements.

| Composition | Band Gap (eV) |
|---|---|
| $Sn_{1.03}Te$ | 0.21 |
| $Sn_{1.01}In_{0.01}Ga_{0.01}Te$ | 0.22 |
| $Sn_{0.99}In_{0.02}Ga_{0.02}Te$ | 0.23 |
| $Sn_{0.95}In_{0.04}Ga_{0.04}Te$ | 0.25 |

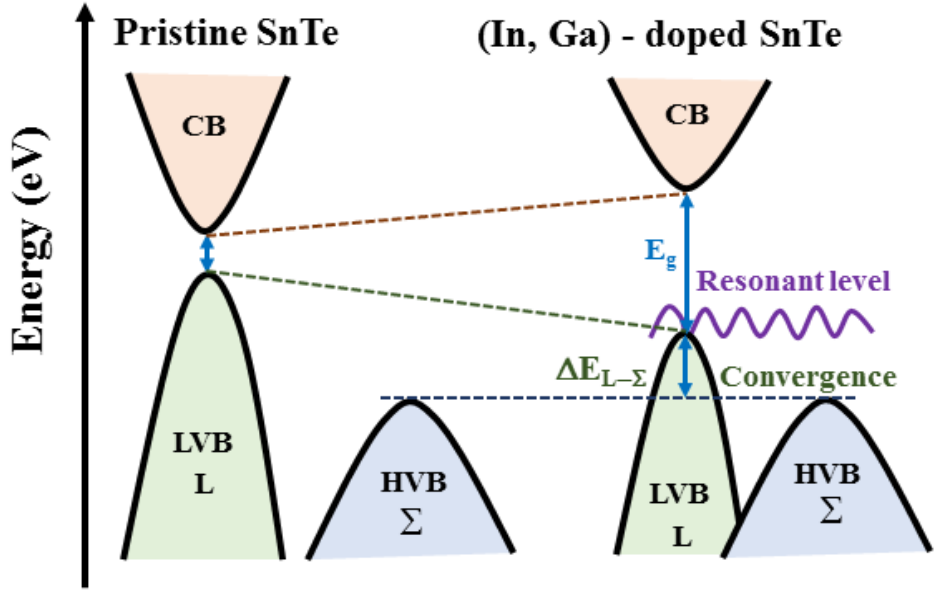

**Figure 9.** Schematic diagram of band structure for pure and In-Ga co-doped SnTe, showing synergistic band convergence and resonant level (Here, CB – conduction band, LVB – light valence band, HVB – heavy valence band).

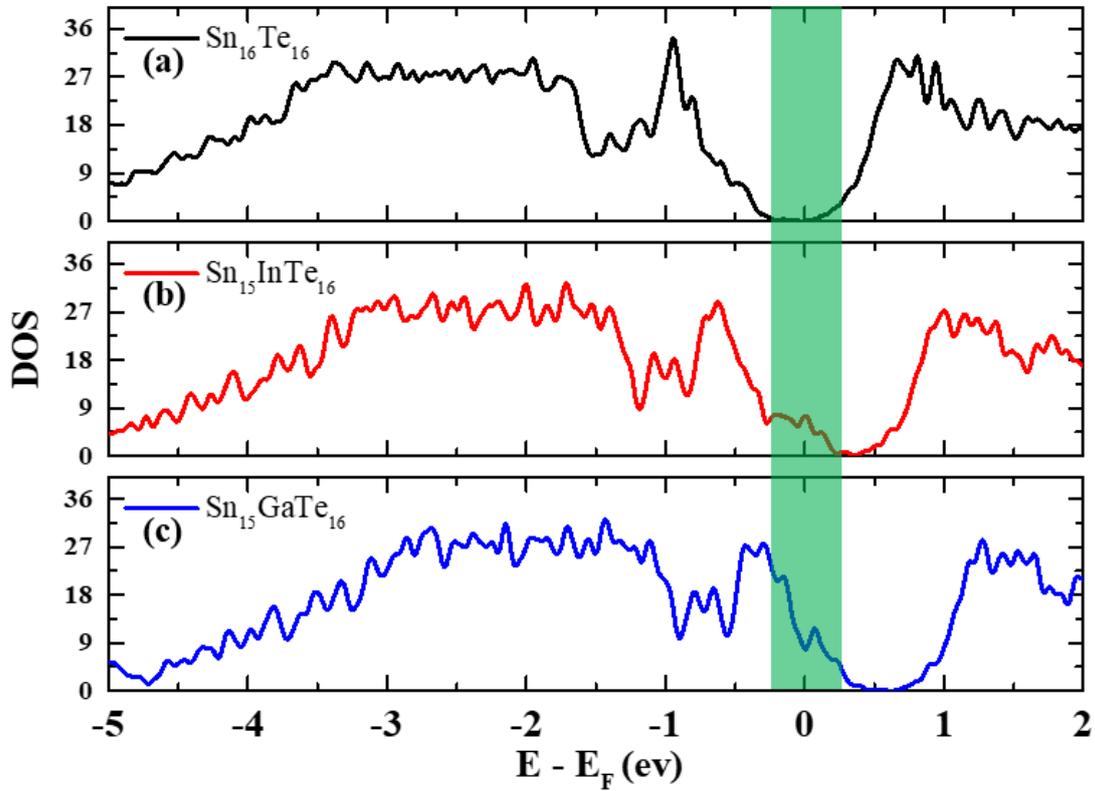

**Figure 10.** Calculated density of states of $Sn_{16}Te_{16}$, $Sn_{15}InTe_{16}$, and $Sn_{15}GaTe_{16}$.

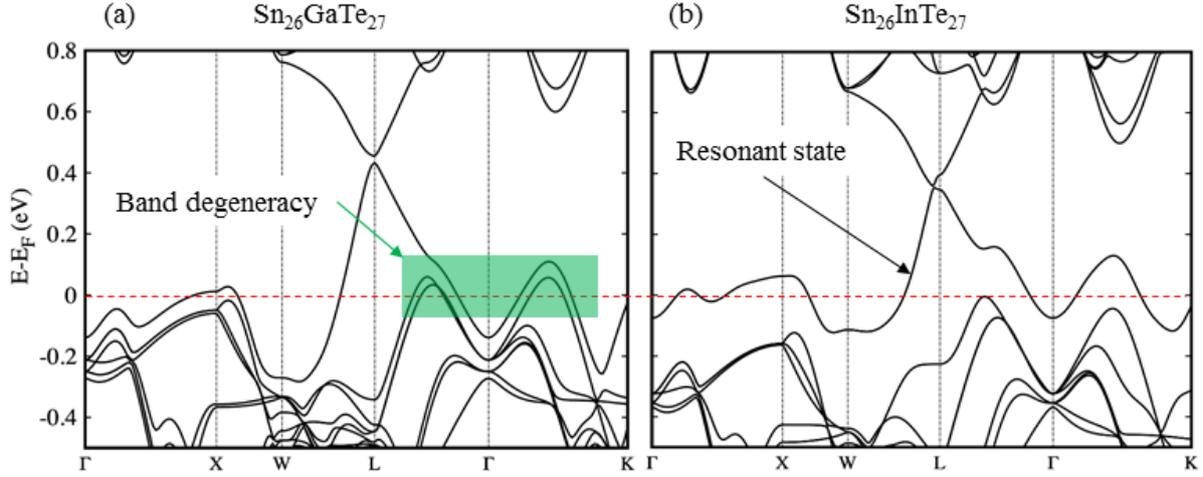

**Figure 11.** Calculated energy band structures $Sn_{26}GaTe_{27}$ and $Sn_{26}InTe_{27}$.

**Computational Results**

Figure 10 shows the total density of states (DOS) of pure SnTe, and the In and Ga doped $Sn_{15}(In,Ga)Te_{16}$ system obtained from first principle calculations. The highlighted region indicates the modification of the DOS modification upon doping. Doping with In and Ga introduces holes at the top of the valence band, leading to a downward shift of the Fermi level from the gap region of undoped SnTe. Figure 11 (a, b) shows the band structure of Ga and In doped SnTe. From Fig 11 (a), it is evident that Ga doping leads to reduction in the energy difference between adjacent energy bands around the Fermi level, leading to band degeneracy. The resonant band, as indicated in Figure 11(b) has contribution mostly from In and Te orbitals. The contribution of individual atoms in the density of states is estimated from their respective partial density of states (p-DOS). Figure S7 (see SM) shows the partial density of states (p-DOS) of elemental orbitals, with a major contribution near the Fermi level in the doped compositions. Indium doping results in a resonant state at the Fermi level due to matching of the In 5s and Te 5p anti-bonding states which can be observed in the p-DOS at the Fermi level (Fig. S7(b)), in accordance with earlier studies [9,52]. On the other hand, the Ga 4s anti-bonding state is not sufficient to form a resonant state with Te 5p (Fig.

S7(c)). The significant increase in the DOS around the Fermi level in the shaded region in $Sn_{15}GaTe_{16}$ compared to $Sn_{15}InTe_{16}$ and $Sn_{16}Te_{16}$ indicates the contribution from Te bands to the observed band degeneracy (Fig. S7(c)), as also reported by previous studies [28,52]. The 5s and 5p orbitals of Sn do not show any significant change with doping. Hence, Sn bands remain mostly rigid and do not participate in the formation of resonant states or in band degeneracy. In the co-doped system, both the factors playing a significant role in transport properties.

**Electrical transport measurements**

The electrical transport properties of pure SnTe and $Sn_{1.03-2x}In_xGa_xTe$ (x = 0, 0.01, 0.02, 0.04) samples were studied in the range 300 – 783 K (Fig. 12(a-d)). The thermal stability of the samples in the above temperature range was confirmed using TGA measurements between 300 – 985 K. The TGA results are shown in Figure S8. The electrical resistivity ($\rho$), shown in Figure 12(a), increases with temperature for all samples, indicative of degenerate semiconductors. Upon In-Ga co-doping, the resistivity shows a non-monotonic behavior, initially decreasing at the lowest doping concentration (x = 0.01), and then increasing as the doping concentration increases. The carrier concentration ($n_p$), and mobility ($\mu$) were determined from room temperature Hall measurements (Fig. 12(d), Table III). The positive sign of the Hall coefficient ($R_H$) indicated p-type conduction in all samples. The carrier concentration of pure SnTe was found to be about $2.6 \times 10^{20}\ cm^{-3}$, which decreased to $9.79 \times 10^{19}\ cm^{-3}$ in $Sn_{1.03}Te$ due to self-compensation of Sn vacancies, in good agreement with previous reports [9,26,28]. The $n_p$ increased on co-doping of In and Ga in the $Sn_{1.03-2x}In_xGa_xTe$ samples, reaching $4.79 \times 10^{20}\ cm^{-3}$ for x = 0.04. The carrier mobility was found to decrease as the doping concentration of In and Ga was increased, probably due to the impurity scattering centers introduced upon doping.

The Seebeck coefficient (*S*) of all samples remains positive and increases linearly in the temperature range 300 - 783 K (Fig. 12(b)), in agreement with the p-type conductivity observed in the Hall effect measurements. The *S* of pure SnTe and self-compensated samples reaches a maximum value of about 80 $\mu$V/K at 783 K, while that for the In-Ga co-doped samples is higher, in the range of 85 - 95 $\mu$V/K at 783 K.

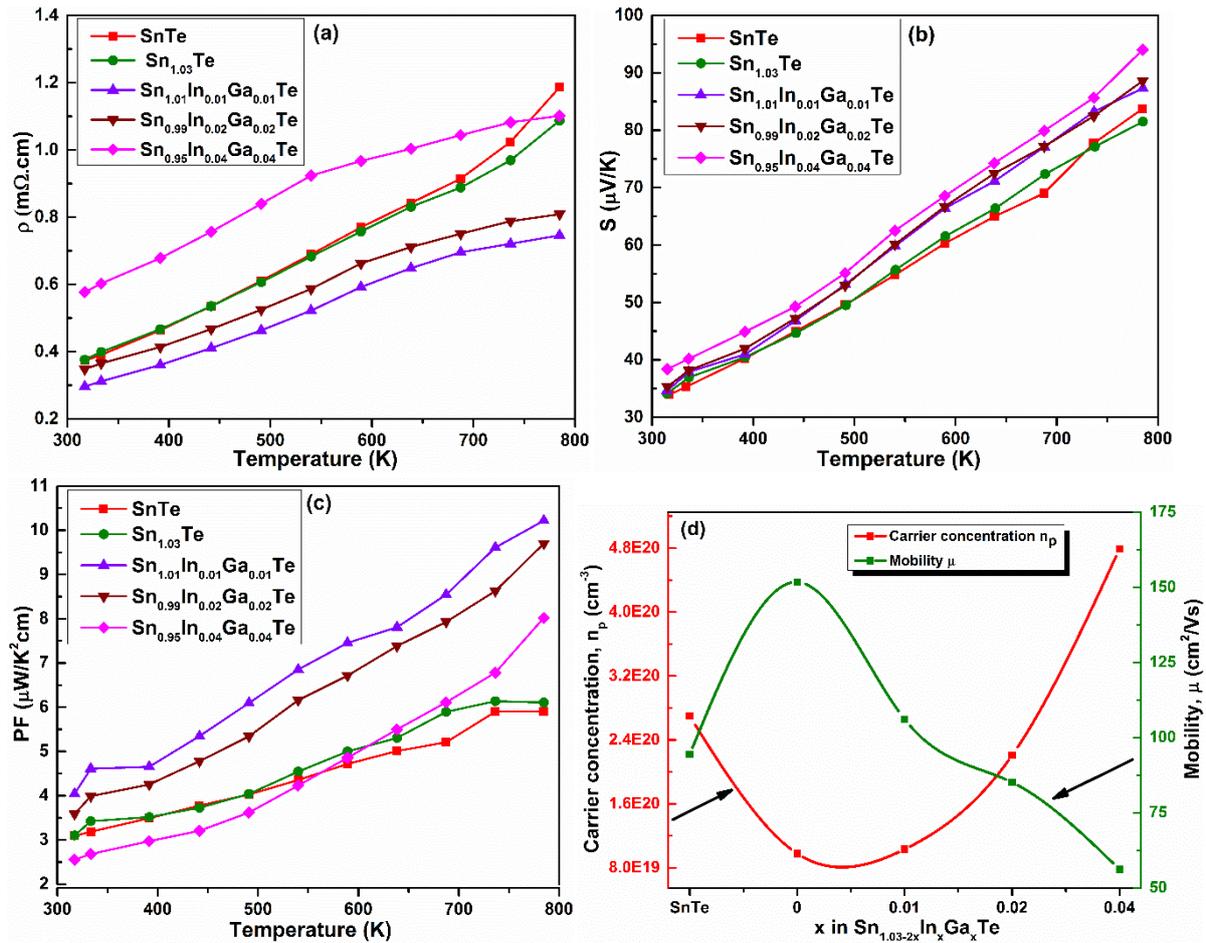

**Figure 12.** (a) Electrical resistivity, (b) Seebeck coefficient, (c) Power Factor, and (d) Room temperature carrier concentration and mobility, of pure SnTe and co-doped $Sn_{1.03-2x}In_xGa_xTe$ (x = 0, 0.01, 0.02, 0.04).

The power factor PF (= $S^2/\rho$), is plotted in Figure 12(c). The (x = 0.01, 0.02) compositions have higher PF, 10.3 and 9.5 $\mu$WK$^{-2}$cm$^{-1}$ respectively at ~ 783 K as compared to pure SnTe (5.8 at 783 K). For x = 0.01 the PF is enhanced by 1.7 compared to pure SnTe. For x = 0.04 the PF is similar to pure SnTe, due to the increased value of electrical resistivity.

The $\rho$ of the self-compensated sample is nearly the same as pure SnTe at room temperature. The temperature variation of the two is very similar with a small divergence above about 650 K. The excess $Sn^{2+}$ leads to a reduction in hole concentration while filling up the Sn site vacancies of pure SnTe. At the same time the carrier mobilities are enhanced, possibly due to reduced scattering from carrier − lattice-vacancy as well as carrier-carrier interactions. As a consequence, the overall resistivity behaviour remains unchanged. Previous studies on SnTe as well as $Sn_{1+x}Te$ doped with hole-donors (Ga, In, Mn, Cd, Ca) have reported an increase in $\rho$ due to scattering from the doped impurities as well as reduced carrier mobility [4,24,27,28,46]. In the present study, the increased hole concentration and scattering from the doped impurities in $Sn_{1.03-2x}In_xGa_xTe$ (specifically the smaller $Ga^{3+}$) can be associated with the observed decrease in $\mu$ and increase in $\rho$ with doping concentration. The initial fall in resistivity for x = 0.01 and 0.02 compared to the pure SnTe and x = 0 samples could be due to a combination of increased carrier concentration and a low concentration of impurity scattering centers.

**Table III.** Carrier concentration ($n_p$), Hall mobility ($\mu$) and effective mass ($m_S^*/m_e$) of pristine SnTe and co-doped $Sn_{1.03-2x}In_xGa_xTe$ (x = 0, 0.01, 0.02, 0.04).

| Composition | $(n_p)\ cm^{-3}$ | ($\mu$) cm$^2$ V$^{-1}$ s$^{-1}$ | $m_S^*/m_e$ |
|---|---|---|---|
| SnTe | $2.69 \times 10^{20}$ | 94.3 | 0.68 |
| $Sn_{1.03}Te$ | $9.79 \times 10^{19}$ | 151.8 | 0.34 |
| $Sn_{1.01}In_{0.01}Ga_{0.01}Te$ | $1.03 \times 10^{20}$ | 106.3 | 0.38 |
| $Sn_{0.99}In_{0.02}Ga_{0.02}Te$ | $2.2 \times 10^{20}$ | 85.2 | 0.63 |
| $Sn_{0.95}In_{0.04}Ga_{0.04}Te$ | $4.79 \times 10^{20}$ | 56.4 | 1.15 |

It has been reported in previous studies that the *S* in doped SnTe is influenced by the formation of resonant levels and by valence band convergence, in the low [9,52] and high [4,24,25,27] regions of the temperature range respectively. Doping with In has been reported to produce both these effects [33,52]. While Ga doping has not been found to introduce resonant levels, it has been reported to produce degenerate hole pockets in the valence band, leading to higher effective mass. Along with increased phonon scattering, this has been reported to enhance the Seebeck coefficient [28]. In the $Sn_{1.03-2x}In_xGa_xTe$ samples of the present study, we observe signatures of the above effects. The increase in optical band gap on doping, determined from FTIR measurements, is indicative of band convergence effects. These results are supported by band structure calculations which show that band degeneracy arises from Ga doping while In doping leads to the formation of resonant levels. We have calculated the effective mass of $Sn_{1.03-2x}In_xGa_xTe$ samples from the measured Seebeck coefficient (*S*) and carrier concentration ($n_p$) at room temperature by using the formulation recently reported by Snyder, et al. [53]. The effective mass increases from $m_S^* \sim 0.68\ m_e$ to $1.15 m_e$ with increasing (In, Ga) doping and is listed in Table III. The increased effective mass is most likely because of the increased contribution of the heavy hole valence band of SnTe [25]. This increased effective mass leads to the observed enhancement in the Seebeck coefficient.

**Thermal transport measurements**

The total thermal conductivity ($\kappa_T$) and lattice thermal conductivity ($\kappa_L$) of pure SnTe, and $Sn_{1.03-2x}In_xGa_xTe$ (x = 0, 0.01, 0.02, 0.04) as function of temperature (300-773 K) is shown in Figure 13(a, b). The total thermal conductivity was calculated using the relation $\kappa = D * C_p * \rho_g$ where *D* is thermal diffusivity, $C_p$ is specific heat, and $\rho$ is density.

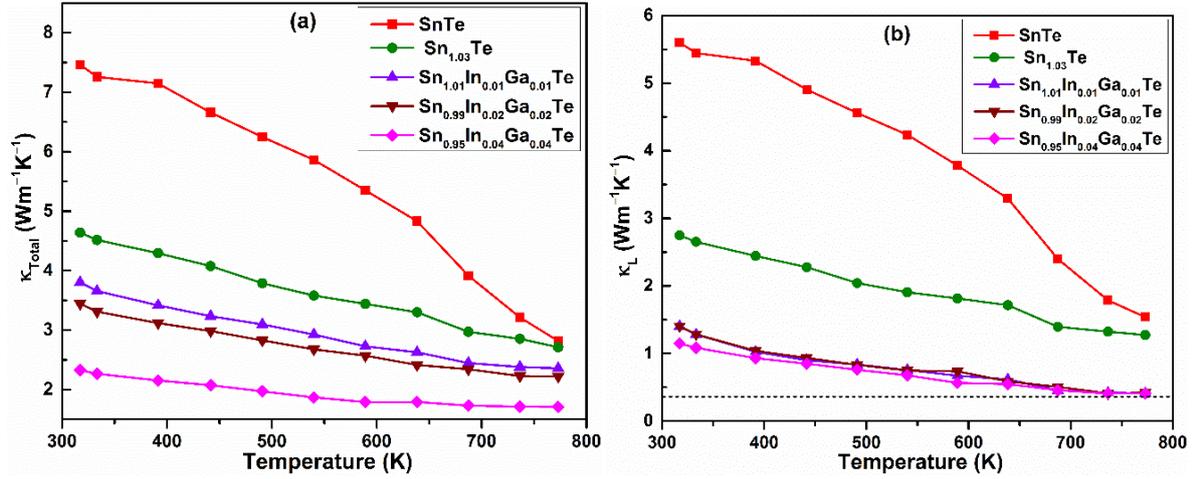

**Figure 13.** Thermal transport properties (a-b): Temperature dependence of (a) total thermal conductivity and (b) lattice thermal conductivity for pure SnTe and co-doped $Sn_{1.03-2x}In_xGa_xTe$ (x = 0, 0.01, 0.02, 0.04).

The electrical thermal conductivity was calculated using the Wiedemann-Franz relation $\kappa_E = LT/\rho$, where the Lorenz number L = 1.5 + exp(-|S|/116), where $S$ is the Seebeck coefficient [54]. The values of L($T$) are plotted in Figure S9. The temperature dependent thermal diffusivities are shown in Figure S10(a). The total thermal conductivity, $\kappa_T$, is found to decrease on doping and decreases with increasing temperature. The value of $\kappa_T$ is reduced by doping from 7.4 W$m^{-1}$K$^{-1}$ to 2.2 W$m^{-1}$K$^{-1}$ at room temperature, and from 2.7 W$m^{-1}$K$^{-1}$ to 1.6 W$m^{-1}$K$^{-1}$ at 773 K. The behavior of the electronic thermal conductivity, $\kappa_E$, shown in Figure S10 (b), is consistent with the variation of the electrical resistivity. Figure 13(b) shows the lattice thermal conductivity, $\kappa_L$, obtained by subtracting $\kappa_E$ from $\kappa_T$. The lattice thermal conductivity of In-Ga co-doped samples is nearly independent of doping concentration over the measured temperature range. The In-Ga co-doped samples exhibited a low value of $\kappa_L$ = 0.42 W$m^{-1}$K$^{-1}$ above 750 K, which is close to the theoretical limit of 0.4 W$m^{-1}$K$^{-1}$ (dotted line) proposed by the Cahill model [55]. This value compares well with those reported in previous studies of co-doped SnTe [28,56,57]. The reduction in the lattice thermal conductivity may originate from a combination of various factors such as scattering of phonons from grain boundaries and point defects [58,59]. This reduction of thermal conductivity contributes significantly towards enhancing the thermoelectric properties of the materials of the present study.

The temperature dependence of the figure of merit $ZT = \frac{S^2 \sigma T}{\kappa}$, calculated from the measured $S$, $\rho$, and $\kappa$ parameters is shown in Figure 14. The $ZT$ values for all samples are found to increase with increasing temperature. The co-doping of In and Ga results in $ZT = 0.35$ at the highest temperature, an increase of 2.1 times that of the pristine compound ($ZT = 0.15$). The $ZT$ values for all doping concentrations are approximately the same at these low concentrations. These values of $ZT$ are comparable to previous studies on SnTe with similar concentrations of dopants [30,32,33].

Some recent studies on heavily doped SnTe with 10 – 15 % dopant have reported $ZT \sim 1$ at high temperatures of about 900 K [28,31,46,50,57]. Thus, the present study on In and Ga co-doping in SnTe can be extended in future by varying the doping concentrations of In (5 – 15 %) and of Ga (2 – 5 %) to optimize thermoelectric performance and $ZT$ in the temperature range of 300 – 923 K.

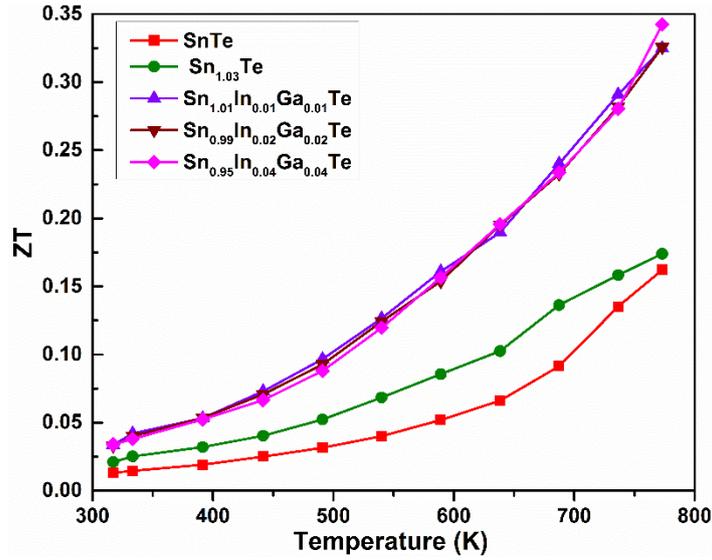

**Figure 14.** Thermoelectric figure of merit, Z*T* as a function of temperature for pure SnTe and co-doped $Sn_{1.03-2x}In_xGa_xTe$ (x = 0, 0.01, 0.02, 0.04).

**Conclusions**

We have investigated the effect of low concentrations of co-doping of In and Ga ions on the thermoelectric performance of SnTe. Samples of pure SnTe and $Sn_{1.03-2x}In_xGa_xTe$ (x = 0, 0.01, 0.02, 0.04) were synthesized by the solid-state route and pelletized using SPS to achieve high density (~99%). All compositions crystallize in a single phase *fcc* (Fm$\bar{3}$m) structure and exhibit a uniform microstructure with homogenous element distribution. The optical band gap was found to increase with In-Ga co-doping, indicative of band convergence effects. First principle electronic structure calculations show formation of degenerate bands near the Fermi level as well as resonant levels at the Fermi level, due to Ga and In doping respectively. Hole-doping by In and Ga led to an increase in carrier concentration but the mobility decreased due to scattering from the dopants. Electrical transport was measured in the temperature range of (300 – 783 K). The resistivity increased with temperature, indicative of the highly degenerate character of the compounds. The Seebeck coefficient of the co-doped samples increased linearly with temperature, reaching 85 - 95 $\mu$V/K at 783 K. At the same time, thermal conductivity decreased sharply with co-doping of In-Ga, and the lattice thermal conductivity above 750 K dropped to 0.42 W$m^{-1}K^{-1}$. A maximum Z$T$ = 0.34 at 773 K was obtained on co-doping, which is about 2.1 times that of the pristine SnTe, due to a combination of the enhanced power factor and low lattice thermal conductivity. The present results suggest that varying the doping concentration of In (x = 0.05 to 0.15) and Ga (x = 0.04) along with techniques such as nano-structural engineering can be used to optimize the thermoelectric properties of SnTe based materials.

ASSOCIATED CONTENT

**Supplemental Material:**

Figures: (S1) Powder XRD pattern; (S2) FullProf Rietveld refinement; (S3) STEM – EDS elemental maps of pure and (In, Ga) doped SnTe; (S4) FESEM micrographs of pristine and $Sn_{1.03-2x}In_xGa_xTe$; (S5) SEM + EDS micrographs of pure and (In, Ga) doped SnTe; (S6) High resolution scans of individual elements present in $Sn_{1.03-2x}In_xGa_xTe$; (S7) Partial DOS of pure and (In, Ga) doped SnTe; (S8) TGA profiles of pure SnTe and $Sn_{1.03-2x}In_xGa_xTe$; (S9) Temperature dependence of Lorenz number; (S10) Temperature dependence of thermal diffusivity and carrier thermal conductivity graphs;

(PDF)

AUTHOR INFORMATION

**Corresponding Author:**

Asad Niazi - Department of Physics, Jamia Millia Islamia, New Delhi – 110025, India.

**Notes:** The authors declare no competing financial interest.ACKNOWLEDGMENT
GJ acknowledges financial assistance from UGC, New Delhi. AK acknowledges the PMRF scheme for student fellowship. MW acknowledges DST, New Delhi for INSPIRE fellowship.28

The authors acknowledge the following for sample synthesis, measurement and characterization facilities: Dept. of Physics and CIF, Jamia Millia Islamia, New Delhi; IUAC, New Delhi; UGC-DAE CSR, Indore; Spark Plasma Sintering user facility, IIT Kanpur; CSIR-NPL, New Delhi; IISER Pune. The authors acknowledge Dr. Prasenjit Ghosh, IISER, Pune and CDAC supercomputing facility (Paramutkarsh), Pune for computational resources. The authors thank Dr. Sunil Nair, IISER, Pune for his help and support in providing the thermoelectric characterization facilities.

.